%% file: main.tex
\documentclass[]{IEEEtran}
\usepackage{latexsym}
\usepackage{graphicx}
\usepackage{amsfonts,amssymb,amsmath}
\usepackage{hyperref}
\usepackage{siunitx}
\usepackage{color}
\usepackage{cite}
\usepackage[dvipsnames]{xcolor}
\usepackage[linesnumbered,ruled,vlined]{algorithm2e}
\newcommand{\G}{\text{G}}
\newcommand{\B}{\text{B}}
\usepackage{tikz}
\usetikzlibrary{automata,arrows,positioning,calc}
\usepackage{pgfplots}
\usepgfplotslibrary{colorbrewer}
\usepackage{bm,verbatim}

\begin{document}

\title{Performance trade-offs in cyber-physical control applications with multi-connectivity
} 

\author{Igor Donevski,\, Israel Leyva Mayorga,\, Jimmy Jessen Nielsen, and Petar Popovski,
       
\thanks{I. Donevski, I. L. Mayorga, J. J. Nielsen, and P. Popovski are with Department of Electronic Systems, Aalborg University, Denmark (e-mail:\{igordonevski, ilm, jjn, petarp\}@es.aau.dk).}
}

\maketitle

\begin{abstract}
Modern communication devices are often equipped with multiple wireless communication interfaces with diverse characteristics. This enables exploiting a form of \emph{multi-connectivity} known as \emph{interface diversity} to provide path diversity with multiple communication interfaces. Interface diversity helps to combat the problems suffered by single-interface systems due to \emph{error bursts} in the link, which are a consequence of temporal correlation in the wireless channel. The length of an error burst is an essential performance indicator for cyber-physical control applications with periodic traffic, as these define the period in which the control link is unavailable. 
However, the available interfaces must be correctly orchestrated to achieve an adequate trade-off between latency, reliability, and energy consumption. This work investigates how the packet error statistics from different interfaces impacts the overall latency-reliability characteristics and explores mechanisms to derive adequate interface diversity policies. For this, we model the optimization problem as a partially observable Markov Decision Process (POMDP), where the state of each interface is determined by a Gilbert-Elliott model whose parameters are estimated based on experimental measurement traces from LTE and \mbox{Wi-Fi}. Our results show that the POMDP approach provides an all-round adaptable solution, whose performance is only $0.1$\% below the absolute upper bound, dictated by the optimal policy under the impractical assumption of full observability.
\end{abstract}


\section{Introduction}
In the rise of the Industry 4.0, the fourth industrial revolution, there is an amassing interest for reliable wireless remote control operations. Moreover, the application of connected robotics, such as in cyber-physical control, is one of the main driver for technological innovation towards the sixth generation of mobile networks \cite{6Gvision}. In accord, one of the main use cases for the fifth generation of mobile networks (5G) is Ultra-Reliable and Low-Latency Communication (URLLC) \cite{popovski2019wireless}. Reliability and latency requirements for this use case are in the order of $1-10^{-5}$ and of a few milliseconds, respectively. The combination of these two conflicting requirements makes URLLC challenging. For instance, hybrid automatic repeat request (HARQ) retransmission mechanisms provide high reliability, but cannot guarantee the stringent latency requirements of URLLC. 
To solve this, recent 3GPP releases have supported dual- and multi-connectivity, in which data packet duplicates are transmitted simultaneously via two or more paths between a user and a number of eNBs. Hereby, reliability can be improved without sacrificing latency by utilizing several links pertaining to the same wireless technology -- 4G or 5G -- but at the cost of wasted time-frequency resources ~\cite{Wolf2019,multiconsurvey}. 
However, modern wireless communication devices, such as smart phones, usually possess numerous wireless interfaces that can be used to  establish an equal number of communication paths. Recent work has proposed \emph{Interface diversity}~\cite{Nielsen2018}, which expands the concept of dual and multi-connectivity to the case where a different technology per interface can be used. Thereby, lower cost connectivity options can help to increase communication reliability.
Since constant packet duplication leads to a large waste of resources, the transmission policies in multi-connectivity and interface diversity systems must be carefully designed to meet the performance requirements while avoiding resource wastage and over-provisioning. Furthermore, as we will observe on the results section, acquiring sufficient knowledge on the channel statistics is essential to attain adequate trade-offs between resource efficiency and reliability in interface diversity systems.

From its definition, the URLLC use case treats each packet individually and, hence, does not capture the performance requirements of numerous applications. For instance, the operation of cyber-physical control applications, that transmit updates of an ongoing process, is usually not affected by individual packets that violate the latency requirements (i.e., untimely packets). Instead, these applications define a \emph{survival time}: the time that the system is able to operate without a required message~\cite{3gppvertical}. Hence, the reliability of communication in such cyber-physical systems is defined by the statistics of consecutive untimely packets, that is, the length of \emph{error bursts}. Hence, in cyber-physical systems, having multiple interfaces with diverse characteristics is greatly valuable, as it allows to select the appropriate interface based on the requirements of the task at hand. For example, while an LTE-based system with multi-connectivity capabilities and selective packet duplication could satisfy the requirements of the application, it seems likely that a combination of unlicensed (e.g. Wi-Fi) and licensed (e.g. LTE) technologies could lead to similar performance guarantees while achieving a lower usage of scarce licensed spectrum and reduce overall costs.

\begin{figure*}[t!]
    \centering
\includegraphics[width=1\linewidth]{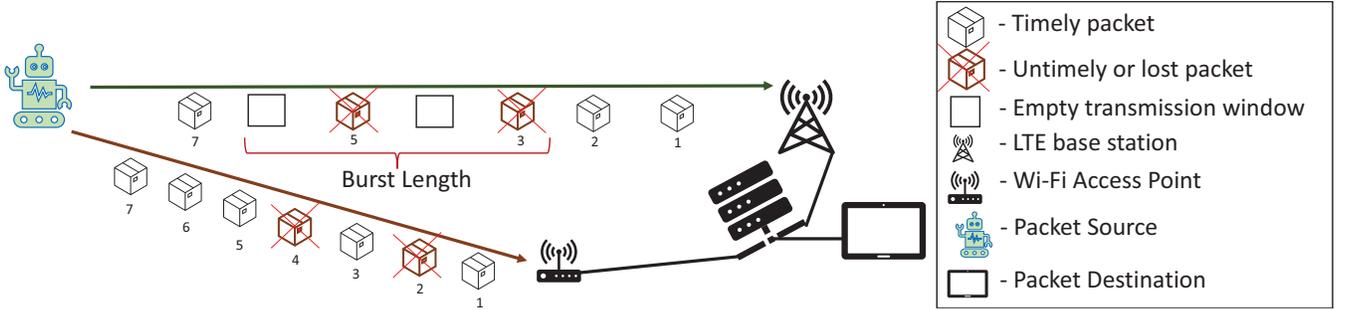}
    \caption{An illustration of the scenario investigating interface diversity where the sender duplicates each packet. The sender would sometimes skip transmission windows in favor of conserving energy. Here, only packet 4 was lost, packets 2,3,5,6 were saved, and packets 1, 7 arrived with a redundant copy.
    }
    \label{fig:illustration}
\end{figure*}
In this paper, we therefore study the performance of interface diversity in terms of burst error distribution in a source-destination system, where we consider two fundamentally different technologies: 1) LTE, which is based on orthogonal frequency division multiple access (OFDMA), operates in licensed spectrum, and where the base station (BS) schedules the uplink resources for communication; and 2) \mbox{Wi-Fi}, which is based on carrier sense multiple access (CSMA)  and operates in unlicensed spectrum. The goal of the proposed interface diversity system, as illustrated in Fig.~\ref{fig:illustration}, is addressing the survival time in cyber-physical control applications \cite{3gppvertical}. In particular,  we investigate the trade-offs between system-lifetime (the time until the system reaches the end of the survival time and operation must be interrupted) and energy consumption. Given the nature of periodic traffic, the survival time can be expressed as the maximum tolerable number of consecutively lost or untimely information packets.

To consider the effect of channel correlation in consecutive errors, we use the Gilbert-Elliott \cite{hasslinger2008gilbert} model that is well suited for representing time-correlated transmissions \cite{GEmeas}. Using this approach, we formulate the problem as a partially-observable Markov decision process (POMDP) that takes into account the limited observability of the inactive interfaces. Hence, based on the observations and the belief states, we can calculate the optimal transmission policy even for devices with extremely limited computational power. We observed that the performance trade-offs  achieved with the POMDP approach are greatly similar when compared the ones achieved with an idealized fully observable MDP. The key contributions of this work are:
\begin{itemize}
  \item The formulation of an interface diversity problem for energy-constrained devices as a POMDP. Hence, our approach considers the limited observability of the inactive interfaces: those that do not transmit, and thus do not receive feedback. While our results are presented for a device using an LTE and a Wi-Fi interface, our model is sufficiently general and, hence, can be applied to cases with more than two interfaces and to different technologies. 
  \item The analysis of interface diversity policies for cyber-physical control applications, where a certain number of untimely packets are tolerated and with error burst due to the temporal correlation in the wireless channel are considered by means of a Gilbert-Elliott model. 
   \item It illustrates that a computationally simple solution, the Q-MDP value method, can be used for solving the POMDP. Using this method we obtain results that closely follow the performance of the fully observable MDP. Specifically, the expected loss in the reward is only around $0.1$\%.
\end{itemize}
  
The rest of the paper is organized as follows. We initially present an elaborate explanation of multi-connectivity and interface diversity's role in timeliness in Section~\ref{sec:sota}. Next, we present the system model in Section~\ref{sec:BEM}, followed by the analysis of the scenario and our proposed method to solve the POMDP in Section~\ref{sec:analysis}. Then, we present the numerical results in Section~\ref{sec:results}. Finally, we conclude the paper with a summary of the work in Section~\ref{sec:conclusion}.

\section{Literature Review}
\label{sec:sota}

Multi-connectivity has been studied from different perspectives. For instance, \cite{Wolf2019} studied a scenario with one user equipment (UE) connected to multiple BSs and with multiple simultaneous connections to the same BS. The benefits of this approach are assessed in terms of transmit power reduction, achieved by increasing the signal-to-noise ratio (SNR). Following a similar multi-connectivity approach, a matching problem is formulated by \cite{Simsek2019}, where the number of UEs in the network and the limited wireless resources are considered. The objective is to provide the desired reliability to numerous users by assigning only the necessary amount of resources to each of them. \cite{Mahmood2018} investigated a similar problem in a heterogeneous network scenario with a small cell and a macro cell. Their results show that multi-connectivity is particularly useful for cell-edge UEs connected to the small cell, and provides even greater benefits when URLLC and enhanced mobile broadband (eMBB) traffic coexist. \cite{8399832} considered multi-connectivity for URLLC as a combination of device-to-device and cellular links, where correlated shadowing is considered. They achieved remarkable increases in the availability ranges for both interfaces. In our previous work~\cite{Nielsen2018} we studied the benefits of interface diversity in terms of reliability for a given error probability. 
Finally, \cite{wifilte} assessed the performance scheduling schemes such as packet duplication and load balancing in order to achieve latency and reliability improvements. The authors exploited a combination of a local Wi-Fi and a private LTE network, that was tested under traffic patterns that are expected to appear in an industrial communications setting.

In the studies mentioned above, only stationary error probabilities are considered. Moreover, \cite{otherPOMDP} provided a thorough investigation of switching off a singular interface that has an unreliable channel, based on channel feedback. The goal of the authors is thus aligned with ours since they aim for an energy-efficient transmission policy given bursty channels, for reliable connectivity of synchronous services. However, the use of different interfaces provides unique benefits for URLLC, especially in the case of bursty wireless errors. For instance, different interfaces are likely to present different burst error distributions, and the correlation of errors between different interfaces is expected to be much lower compared to the correlation between multiple links using the same wireless interface. Despite these evident benefits, and the thorough investigation of burst errors in past research \cite{Yajnik}, little research has been conducted on interface diversity with error bursts. Specifically, our previous work presents one of the few analyses of this kind~\cite{Nielsen2019}. However, it was limited to the benefits of interface diversity in the length of error and success bursts without considering the impact on resource efficiency.

Cyber-physical control applications can belong to one of two major categories depending on the traffic direction requirements: in downlink or uplink only (open-loop control) or the combined uplink and downlink (closed-loop control) requirements~\cite{3gppvertical}. Moreover, a closed-loop control application needs to process incoming events, and thus give appropriate instruction commands to those events \cite{closedloopOLD}. In such scenarios, timeliness is critical to avoid violating the system imposed latency requirements, which leads to executing outdated actions. Therefore, being untimely is the equivalent of a failure in communication service availability. 

Open-loop control applications with periodic commands appear frequently in industrial applications and are considered representative of cyber-physical control systems. In these applications, failing a specific number of consecutive updates directly corresponds to exceeding the survival time and, hence, to an error in the system. For example, it has been observed that the number of consecutive errors impacts the stability of the system and leads to a considerable decrease in safety of autonomous guided vehicles~\cite{DeSantAna2020}. As in the present model and in our previous work~\cite{Nielsen2019}, a Gilbert-Elliott model was considered in~\cite{DeSantAna2020} to introduce correlation in the wireless channel. Finally, the novelty of this work comes from investigating the problem of interface diversity for timely packet arrivals for cyber-physical control applications in a burst error channel, where the reliability of the system comes as a trade-off of energy.

\section{System model}
\label{sec:BEM}
We consider a point-to-point communication between a user and a BS in an industrial scenario. The user samples a given set of physical phenomena and generates data periodically, where $T_s$ is the sampling period. The sampled data is immediately transmitted to the BS, where it is used for control purposes, so that it must be received within a pre-defined latency constraint $\theta\leq T_s$. Hence, it is now convenient to introduce the definition of the latency-reliability function, which stands for the probability of being able to transmit a data packet from a source to a destination with a given latency deadline~\cite{Nielsen2018}.

Let $L$ be the RV that defines the packet latency. Then, for a given interface $i$ and latency deadline $\theta$, the latency-reliability function is defined as
\begin{equation}
    F_i\left(\theta\right) = \Pr\left(L\leq \theta\mid i\right).
\end{equation}
As such, the latency-reliability function is a CDF of the interface's latency, where lost packets have the equivalent of infinite latency. 
Thus, the error probability becomes a specific value (of deadline) $\Theta$ in the latency-reliability function, and we define the probability of error for interface $i$ as
\begin{equation}
    P_e^{(i)}=1- F_i\left(\theta\right).
\end{equation}
It should be noted that the traditional definition of the probability of error is obtained for the case $\theta\rightarrow \infty$ and that the distribution of $L$ can be updated continuously to reflect the changes in the wireless channel.

We consider the case where the interface diversity system uses packet cloning, where a full packet is transmitted via each of the $N$ interfaces. Next, by assuming that errors across the multiple available interfaces occur independently, the end-to-end error probability can be calculated as in \cite{billinton1992reliability,Nielsen2018}:
\begin{equation}
    P_e^\text{E2E} = \prod_{i=1}^N (1 - F_i\left(\theta\right)) = \prod_{i=1}^N P_e^{(i)}.
\end{equation}
Note that the correlation of the large-scale fading across the interfaces is captured by the model described above through the distribution of the RV $L$. The assumption of errors occurring independently across interfaces holds since correlation in the fast fading may only occur if the antenna elements within an array have insufficient spacing and/or if the concurrent transmissions occur in frequencies that are separated by less than one coherence bandwidth \cite{correlation}. In out case, the use of two different technologies and frequency bands for WiFi (unlicensed ISM bands) and LTE (licensed spectrum) ensure that the transmissions are sufficiently separated in frequency to avoid correlation.

The BS sends individual feedback per interface to the user after each transmission attempt. 
If the data is not received within $\theta$, it is declared as missing and the user receives a NACK. The system tolerates a maximum number of missed transmissions. Specifically, if the number of missed transmissions is $N$, the system declares a failure and operation must be interrupted. Otherwise, the system is able to continue normal operation whenever the number of missed transmissions is $n\leq N$ (smaller than the survival time). 

In the following, we define our interface diversity problem as a POMDP denoted as the tuple $\left(\mathcal{S}, \mathcal{A}, T, \mathcal{R}, \Omega, \mathcal{O}\right)$. Here, $\mathcal{S}$ is the set of states, $\mathcal{A}$ is the set of actions, $T$ is the transition probability to the next state of the \emph{environment} given a state-action pair, $\mathcal{R}\subset\mathbb{R}$ is the set of immediate rewards, $\Omega$ is the set of possible observations, and $\mathcal{O}$ is the observation probability when transitioning.
\begin{figure*}[t]
    \centering
    \input{figures/gilbert_elliott_4state}
    \caption{(a) Two-state GE model for interface $i$ and (b) four-state GE model for user with two interfaces.}
    \label{fig:ge_model}
\end{figure*}
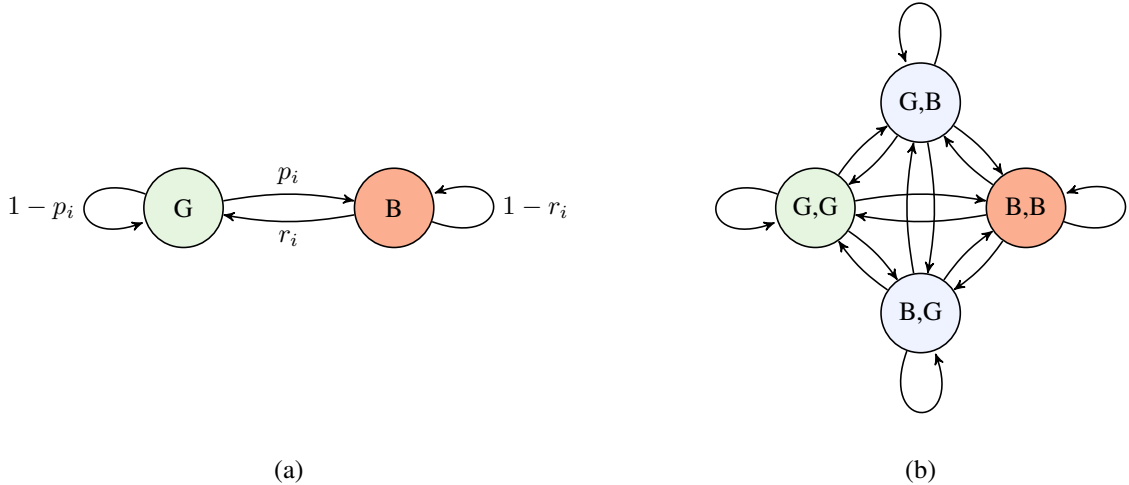

\subsection{The environment}
The user, i.e. the agent, interacts with the \emph{environment} at discrete time steps $t\in\mathbb{N}$ by sampling and transmitting data to the BS. It is equipped with two distinct communication interfaces, $i\in\{1,2\}$, where the generated data can be transmitted. Throughout this paper, we assume that these interfaces are completely independent from each other and that each interface is accurately modeled by a two-state Gilbert-Elliott (GE) model \cite{hasslinger2008gilbert}. The GE model was selected due to its simplicity and to the ability to capture temporal correlation.

In our GE model, the details of the implemented protocol and the wireless conditions -- interference, noise, and fading -- are simplified and related to two possible states in a discrete-time Markov chain (DTMC). These are the \emph{good} state G and the \emph{bad} state B. Hence, we model the state space for the GE model for interface $i$ as $\mathcal{S}_i=\{\G,\B\}$. At any given point in time, an interface is in the G state if the protocol and the wireless channel conditions are such that allow for a transmission to be received within the latency constraint $\theta$.  Otherwise, the interface is in the B state. 

This simple GE model has two parameters, namely $p_i$ and $r_i$ that determine the transition probabilities and, hence, the steady-state error probability and burst lengths~\cite{hasslinger2008gilbert}. Hence, these are system- and environment-specific, and can only be learned after deployment by collecting statistics of the packet transmissions. An additional benefit of using the GE model is that through continuous tuning of the statistical parameters, it allows to capture cross-interface correlation due to large-scale fading or traffic surges.
 
We denote the state of interface $i$ at time step $t-1$ as $s_i$ and as $s'_i$ at time step $t$. Parameter $p_i$ represents a transition from state G to B and $r_i$ from B to G (i.e., a recovery from the bad state). Hence, the transition probabilities are defined as 
  \begin{align}
    \Pr\left(s_i'=\G\mid s_i=\G\right)&=1-p_i,\\
    \Pr\left(s_i'=\G\mid s_i=\B\right)&=r_i,\\
    \Pr\left(s_i'=\B\mid s_i=\G\right)&=p_i,\\
    \Pr\left(s_i'=\B\mid s_i=\B\right)&=1-r_i,
\end{align} 

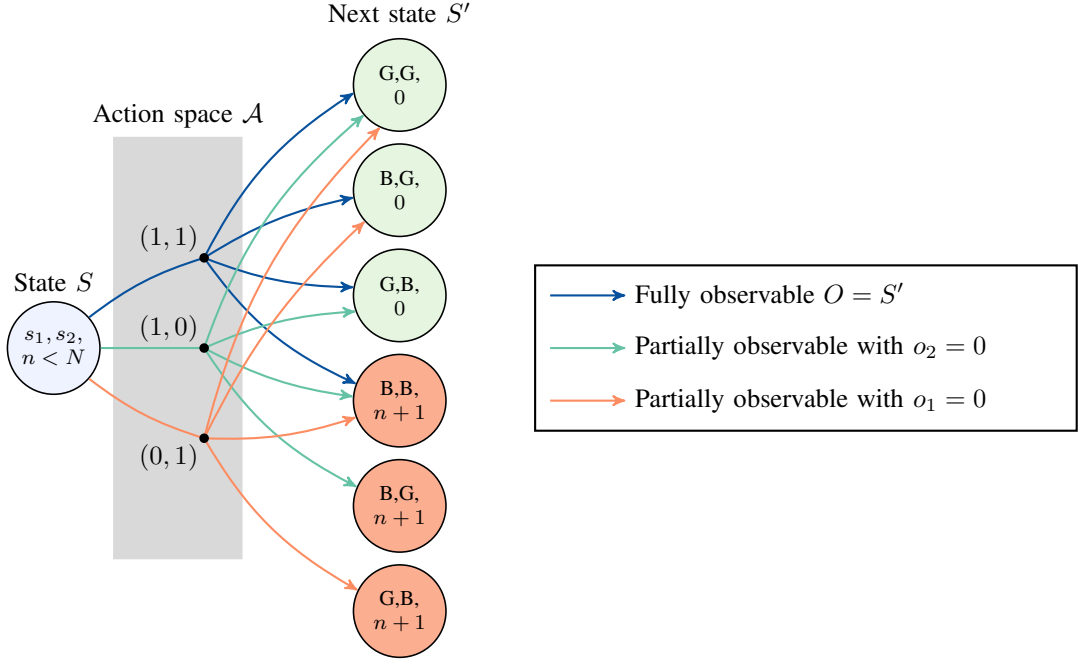
\begin{figure*}[t]
    \centering
    \input{figures/pomdp}
    \caption{One-step transitions from an arbitrary non-absorbing state (i.e., $n<N$) for the partially observable Markov decision process (POMDP) with two interfaces.}
    \label{fig:pomdp}
\end{figure*}
Fig.~\ref{fig:ge_model}a illustrates the GE model with one interface, whose transition probability in a matrix form is:
\begin{equation}
\mathbf{P}_i = 
 	\begin{bmatrix}
    1-p_i & p_i \\
    r_i& 1-r_i
  \end{bmatrix}.
  \label{eq:GEmatrix}
  \end{equation}

Building on this, the state of a system with two interfaces is defined by the four-state GE model illustrated in Fig.~\ref{fig:ge_model}b, where transition labels are omitted for brevity. To elaborate, transition probabilities are calculated under the assumption of the two interfaces being independent, for example, the transition from state G,G to state B,B has probability $p_1 p_2$.
  
Besides the status of each interface, knowing the number of consecutive missed data $n$ is essential for the operation of the system. 
Therefore, to build a respective Markov decision process (MDP), we define the state space as $\mathcal{S}=\{\left(s_1,s_2,n\right)\}$, where  $n\in\{0,1,2,\dotsc,N\}$. Note that the true state of the interface is observable only after concluding the transmission at time step $t-1$, and before the following transmission attempt at time $t$. Thus, the state of the environment at time $t$ denoted as $S\in\mathcal{S}$ is defined by the outcome of the transmission in the last attempt at $t-1$. Hence, all states $S\in\mathcal{S}:n=0$ indicate that a transmission at time $t-1$ was received successfully. Furthermore, all states $S\in\mathcal{S}:n=N$ are absorbing states and, consequently, this is a finite MDP with episodic tasks. 

Ultimately, the goal of the system is to navigate the MDP in a way that decreases the amount of errors, and altogether reduces the likelihood of having $N$ consecutive errors. Thus, the system should be incentivised to maximize its expected lifetime, while optimizing the costs associated with each transmission. This is done through proper allocation of rewards for each action in the MDP. However, the main challenge for solving the issue comes as a product of the limited observability of the defined MDP system when an interface is switched off. The details for this are encompassed in the following subsection.

\subsection{Actions, rewards, and uncertainty}
At each time step $t$ (i.e., data transmission instant) the user takes an action $A\in\mathcal{A}=\{(a_1,a_2)\}$, where $a_i\in\{0,1\}$; $a_i=1$ indicates transmission and $a_i=0$ indicates no transmission for interface $i$. Hereafter we denote that interface $i$ is on when $a_i=1$ and off otherwise. Note that in our case $\mathcal{A}\left(S\right)=\mathcal{A}$ for all $S\in\mathcal{S}$; that is, the set of actions is the same in every possible state. This totals to three different actions -- leaving out the option to turn off all interfaces altogether $A=(0,0)$ -- and have either interface off, or both interfaces turned on. 

Having taken action $A$ when in state $S$ there is a probability $T(S,A,S')$ to end up in state $S'$, therefore it must apply that $\sum_{S'}T(S,A,S')=1$. Note that $S'$ represents the true state of both interfaces at time $t$, that is revealed only after taking action $A$. A missed transmission can thus occur when both interfaces are transmitting $A=(1,1)$ but are in the bad state $S'=(B,B,n)$, or a single interface is transmitting that is in its corresponding bad state -- $A=(0,1)$ when $S'=(G,B,n)$, or $A=(1,0)$ when $S'=(B,G,n)$.

Therefore, following a transmission action $A$, the user receives a reward $R$ that also depends on the feedback by the BS given before time $t+1$. Having arrived at state $S'$ by taking action $A$ when in state $S$, yields a reward $R(S,A,S')$ that is a function $r(n)$, where $r(n=0)=1$ is a successful transmission and $r(n>0)=-1$ is a missed transmission. In the overall reward allocation, we also account for the cost of using interface $i$, specifically,
 \begin{equation}
     R(S,A,S') =r(n)-c(A)=r(n)-a_{1}c_1-a_{2}c_2,
 \end{equation}
 where $c(A)$ is the cost of taking action $A$.
 
 In an MDP, a policy $\bm{\pi}$ is a function that maps each state $S\in\mathcal{S}$ to an action $A\in\mathcal{A}(s)$. Therefore, given a policy $\bm{\pi}(S)$, the agent will choose action $A$, once it finds itself in state $S$. 
  Our objective is to find an optimal policy $\bm{\pi}^*$ that selects the best action given some state $S$. Let $K$ denote the total number of time steps until the system transitions to an absorbing state. The best action at time $t$ is the one that maximizes the discounted return 
  \begin{equation}
      G_t=\sum_{k=0}^{K-t-1}\gamma^k R_{t+k+1},
  \end{equation}
where $0<\gamma<1$ is the discount factor, and the system lifetime is $K<\infty$ as there is always a non-zero probability of transitioning to the absorbing states in a finite number of steps. Thus, the value of the MDP process, when starting from initial state $S_\text{init}$ under a policy $\bm{\pi}$, is:
\begin{equation}
    V_\pi(s)=\mathbb{E}_\pi\left[G\middle|S = S_\text{init}\right].
\end{equation}

Next, we define the set of observations as $\Omega=\{\left(o_1,o_2,n\right)\}$, where $o_i\in\mathcal{S}_i\cup\{0\}$ is the set of observations for an interface $i$, and the number of consecutive missed deadlines $n$ is always observable. The transition from $S$ to $S'$ following action $A$ provides a deterministic observation $O=(o_{1},o_{2},n)\in\Omega$ in the following manner: (Since all observations are deterministic with each action, the set  $\mathcal{O}$ is irrelevant to the analysis.)
I) Having an interface on, namely $a_i=1$, allows for fully observing the state of that interface $o_i=s_i$. II) On the other hand, having an interface off, namely $a_i=0$, provides no observation of the state of that interface, i.e. $o_i=0$, unless additional mechanisms are available to correctly estimate the state, for example, based on the exchange of control messages. Building on this, we define the following function.
\begin{equation}
    o_{i} = \begin{cases} s_{i} & \text{if } a_{i}=1 \textsc{ or } \alpha=1,\\
    0 & \text{otherwise},
    \end{cases}
\end{equation}
where $\alpha=1$ indicates that the BS has additional mechanisms to perform the observation.
Fig.~\ref{fig:pomdp} illustrates the action space along with the associated transitions to states $S'$ and observations from an arbitrary non-absorbing state $S$.

\section{Analysis}
\label{sec:analysis}
As a first step, we derive the steady-state probabilities of the good and bad states from the transition matrix $\mathbf{P}_i$ as \cite{hasslinger2008gilbert}:
\begin{align}
	\pi_{i,\G} = \frac{r_i}{p_i+r_i} \\
	\pi_{i,\B} = \frac{p_i}{r_i+p_i} \label{eq:PiB}
\end{align}
where $\pi_{i,\G} + \pi_{i,\B} = 1$.

Assuming that the system has both interfaces turned on during initialization, the initial state of the MDP $S_\text{init}$ is chosen randomly among the set of states $\{ (G,G,0), (G,B,0), (B,G,0), (B,B,1) \}$ based on the steady state probabilities $\pi_{i,\G}$ and $\pi_{i,\B}$.

\subsection{Policy Utility Through the Value and Q functions}
Let $V_\pi(S)$ be the expected utility received by following policy $\pi$ from state $S$, as in:
\begin{equation}
    V_\pi(S) = \begin{cases} Q_\pi(S,A) & \text{if } S \notin \{(s_1,s_2,N)\},\\
    0 & \text{otherwise},
    \end{cases}
\end{equation}
where $Q_\pi(S,A)$ is the expected utility of taking action $A$ from state $S$ , and then following policy $\pi$ \cite{puterman2014markov}. Starting from state $S_\text{init}$ our goal is to find the optimal policy $\pi^*$ that results in the maximum value that can be obtained through any policy $V_{\pi^*}(S_\text{init})$. As illustrated in Fig. \ref{fig:vandq}, when not following the optimal policy $\pi^*$, but sampling the value of the Q-functions for each action $A$ in the space of values $V_{\pi^*}(S')$, we get:
\begin{figure}[t!]
    \centering
\includegraphics[width=1\columnwidth]{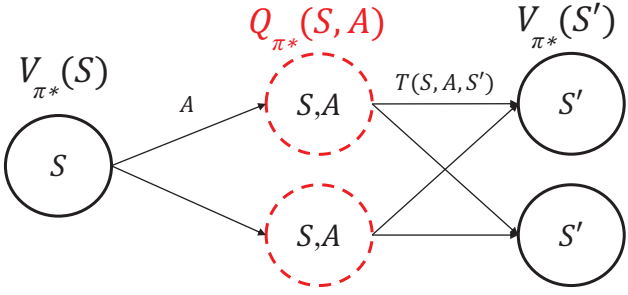}
    \caption{An illustration of a single step of the value iteration process for state $S$ that tests the Q-function for all state-action pairs, where future states follow the optimal policy $\pi^*$}
    \label{fig:vandq}
\end{figure}
\begin{equation}
    Q_{\pi^*}(S,A) = \sum_{S'}T(S,A,S')[R(S,A,S')+\gamma V_{\pi^*}(S')],
\end{equation}
where, $\gamma$ is the discount factor that controls the importance of short term rewards ($\gamma$ values close to 0), or long term rewards ($\gamma$ values close to 1), where the anticipated rewards are represented through the value of the state as:
\begin{equation}
    V_{\pi^*}(S) = \begin{cases} \max\limits_{A \in \mathcal{A}} Q_{\pi^*}(S,A) & \text{if } S \notin \{(s_1,s_2,N)\},\\
    0 & \text{otherwise},
    \end{cases}
\end{equation}
Unless the user is in an $N$ state that is absorbing, the values $V_{\pi^*}(S)$ are recurring and can be approximated through the iterative process of \cite{puterman2014markov}:
\begin{equation}
    V_{\pi^*}^k(S) \longleftarrow \max_{A \in \mathcal{A}} \sum_{S'}T(S,A,S')[R(S,A,S')+\gamma V_{\pi^*}^{k-1}(S')],
\end{equation}
that continues until it converges to some predefined precision $\epsilon$ of the past and current value:
\begin{equation}
    \max_{S \in \mathcal{S}} |V_\pi(S)^{k} - V_\pi(S)^{k-1}| \leq \epsilon.
\end{equation}
Thus the method of value iteration guarantees finding the optimal value for an MDP that is a function of the optimal policy. Given a $\kappa$ number of iterations to converge to a solution, the complexity of this algorithm is $O(\kappa SAS')$, which given our small MDP, is insignificant. Unfortunately, this method does not directly produce an optimal policy for a POMDP, however, this can be addressed by the $Q_\text{MDP}$ value method.

\subsection{Belief Averaged $Q_\text{MDP}$ Value Method}
\label{sec:optpol}
Due to the limited information on the channel properties for each interface our Markov process is a POMDP where we cannot fully observe the true state space for time $t$.
Therefore, the agent maintains a belief $b$ on the state of the system $S\in\mathcal{S}$ based on the observation $O\in\Omega$. By observing that there is no uncertainty on the value of $n$ and that the state of the interfaces $\mathcal{S}_i$ is independent, we can define the belief as:
\begin{multline}
            b(S,O)=\Pr\left(S=(s_1,s_2,n)\mid O=(o_1,o_2,n)\right)  \\  = b_1(s_1,o_1) b_2(s_2,o_2),
\end{multline}
where $b_i$ is the belief for interface $i$ to be in state $s_i$, given its observation $o_i$. The $b_i$ values are updated recursively with each following observation as in:
\begin{equation}
   b_i(s_i,o_i)  \longleftarrow \Pr\left(s_i\mid o_i\right) = \begin{cases}       1 & \text{if } s_i=o_i \text{ and  } o_i \neq 0 \\
                                                    f_{s_i}(b_i) & \text{if } o_i = 0,\\
                                                    0  & \text{otherwise},
    \end{cases}
\end{equation}
where $f_{s_i}(b_i)$ is a function for calculating the probability of being in state $s_i$ as a function of the previously held beliefs in: 
\begin{align}
   f_{G}(b_i) =& (1-p_i) b_i(G,0) + r_i b_i(B,0),   \\
   f_{B}(b_i) =&p_i b_i(G,0) + (1-r_i) b_i(B,0).
\end{align}
Therefore, knowing the belief $b(S,O) \forall S \in \mathcal{S}$ we can proceed with finding an optimal policy for the underlying POMDP through the following two steps.

Step 1: Ignore the observation model and compute the Q-values $Q_{\pi^*}(S,A)$ given directly from the state-action pairs. These are denoted as $Q_\text{MDP}\left(S,A\right)$ and are obtained through calculating the Bellman operator in the value iteration method.

Step 2: Calculate the belief averaged Q-values for each action and belief $b(S,O)$ as:
\begin{equation}
    Q_A(b)=\sum_{S\in\mathcal{S}}b(S,O)\, Q_\text{MDP}\left(S,A\right).
\end{equation}
The optimal policy now becomes a function of the belief, instead of the current state, and is
\begin{equation}
    \pi^*(b) = \max_{A \in \mathcal{A}}Q_A(b).
\end{equation}
Note that this is a method that does not incentivize updating the belief state, but optimizes with the assumption that we will have full observability following the transmission at time $t$ \cite{LITTMAN1995362}. 

\subsection{Parameter Tuning}
For the MDP to optimize the operation of the underlying communications system we require a proper assignment of the rewards and costs for the MDP. Therefore, the reward and punishment for a successful or a missed transmission were fixed to $1$ and $-1$, respectively. Conversely, the value of $c_i$, the cost of using interface $i$, greatly depends on the specific characteristics of the system and on the individual notion of resource efficiency. Moreover, the cost of using an interface is directly related to the consumption of resources that would otherwise be available to other services. Throughout the rest of the paper, we consider that the cost of using an interface is given by the energy consumption. However, other parameters can be used to define the cost of each interface when adapting our methods to a specific system.

Given a transmission power $E_\text{LTE}$ and $E_\text{Wi-Fi}$ for the LTE and the Wi-Fi interface, respectively, we calculate the cost for interface $i$ as
\begin{equation}
c_i = \eta\frac{E_i}{\sum_i E_i},
\end{equation}
where $\eta$ is a cost scaling factor that serves to reduce/increase the importance of the energy transmission costs with regards to the initial rewards. The scaling factor $\eta$ was sampled across several values in the range $0 \leq \eta \leq 1$, which resulted in five different optimal policies, one for each different $\eta\in\{ 0, 0.03, 0.07, 0.2, 1\}$.
\subsection{Latency measurements for modeling Wi-Fi and LTE}
\label{sec:testbed}
\begin{figure}[tb]
	\centering
	\includegraphics[width=1\linewidth]{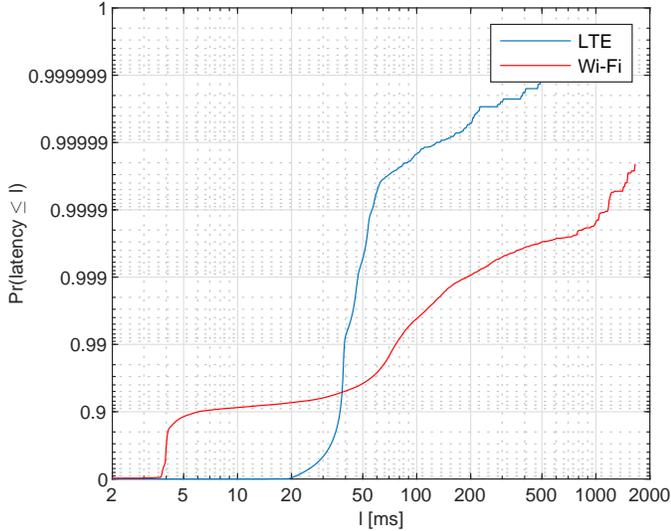}
	\caption{Empirical latency CDFs of considered interfaces.}
	\label{fig:empirical_lcdf}
\end{figure}
Traces of latency measurements for different communication technologies were obtained by sending small (128~bytes) UDP packets every 100~ms between a pair of GPS time-synchronized devices through the considered interface (LTE, or \mbox{Wi-Fi}) during the course of a few work days at Aalborg University campus. 
A statistical perspective of this data is given by the latency CDFs in Fig.~\ref{fig:empirical_lcdf}, which clearly outlines some key differences between the performance of the LTE and \mbox{Wi-Fi} interfaces. While \mbox{Wi-Fi} can achieve down to 5~ms one-way uplink latency for 90\% of packets, it needs approx. 80 ms to guarantee delivery of 99\% of packets. For LTE, on the other hand, there is hardly any difference between the latency of 90\% and 99\% delivery rates, approx. 36~ms and 40~ms, respectively.
Since the measurements for both LTE and \mbox{Wi-Fi} were recorded in good high-SNR radio conditions, we expect that the differences between LTE and \mbox{Wi-Fi} can, to a large extent, be attributed to the inherent differences in the protocol operation and the fact that LTE operates in licensed spectrum whereas \mbox{Wi-Fi} has to contend for spectrum access in the unlicensed spectrum.

\begin{table*}[t]
\caption{Parameters for evaluation}
\begin{center}
\renewcommand{\arraystretch}{1.3}
\begin{tabular}{lll}
\hline
Label & Definition & Value \\ \hline
\hline
$\theta$ & Latency Constraint & 38.25 ms \\
$p_\text{LTE}$ & LTE's p-transition probability & 0.0178 \\
$p_\text{Wi-Fi}$ & Wi-Fi's p-transition probability & 0.0515\\
$r_\text{LTE}$ & LTE's r-transition probability & 0.2577 \\
$r_\text{Wi-Fi}$ & Wi-Fi's r-transition probability & 0.9468 \\
$E_\text{Wi-Fi}$ & Power consumption of Wi-Fi interface & 15.85 mW \\
$E_\text{LTE}$ & Power consumption of LTE interface & 200 mW \\
$N$ & Maximum number of consecutive missed transmissions & 4  \\
$\epsilon$ & Value iteration convergence criteria & $10^{-11}$  \\
$k_\text{max}$ & Value iteration maximum number of iterations & $10^{5}$  \\
$\gamma$ & Discount factor & $0.99999$  \\
\hline
\end{tabular}
\end{center}
\label{tab:setting}
\end{table*}

\subsection{Performance evaluation}
To conduct the performance evaluation of the policies obtained with the POMDP, we define the following benchmarks. 
\begin{itemize}
    \item \emph{Fully observable system:} Assumes an inherent ability of the BS to inform the user about the interface that is turned off, for example, by using pilots that precede the transmissions. In this case $\alpha=1$, making the POMDP collapse to an MDP. We denote the policy with full observability as $\bm{\pi}^*_{\alpha=1}$.
    \item \emph{Forgetful POMDP (F-POMDP):} Maintains a single state of partial belief and, afterwards, assumes the steady state probability $\pi_{i,\G}, \, \pi_{i,\B}$ for the inactive interface. This forgetful approach collapses to a small MDP where belief does not need to be continuously computed. 
    \item \emph{Hidden MDP (H-MDP):} Is the fully reduced MDP of the forgetful approach, where the belief averages in F-POMDP are joint in a single state. Here, the transition probabilities for the inactive interface directly become the steady state probabilities $\pi_{i,\G}, \, \pi_{i,\B}$.
\end{itemize} 

The obtained policies are evaluated based on the following performance indicators. First, the distribution of the number of consecutive errors $n$. Second, the utilization of the LTE interface, defined as the ratio of time slots when the LTE interface is turned on $u_\text{LTE}(\bm{\pi})=\Pr\left(a_1=1\mid \bm{\pi}\right)$. Third, the expected system-lifetime, defined as the number of time steps from initialization until the system transitions into an absorbing state. For the latter, let $\overline{K}(\bm{\pi})$ be the expected system lifetime with policy $\bm{\pi}$. Finally, we define the expected total reward of the system with policy $\bm{\pi}$, from initialization until absorption, as $\overline{R}(\bm{\pi})$.

Building on this, we assess the policies derived with partial observability w.r.t. the policy with full observability based on:
\begin{itemize}
\item \emph{System lifetime delta:} Denotes the relative increase of the expected system lifetime w.r.t. the MDP with $\alpha=1$ (i.e., full observability), denoted as:
\begin{equation}
   \Delta K =\frac{\left(\overline{K}(\bm{\pi})-\overline{K}(\bm{\pi}^*_{\alpha=1})\right)}{\overline{K}(\bm{\pi}^*_{\alpha=1})}.
\end{equation}
Hence, positive values of $\Delta K$ indicate an increase in the system lifetime w.r.t. the optimal policy with full observability.
\item \emph{Policy deviation:} Measures the relative change in the expected system lifetime $K$ and expected transmission cost as 
\begin{equation}
    \Delta \bm{\pi}=|\Delta K|+{|u_\text{LTE}(\bm{\pi})-u_\text{LTE}(\bm{\pi}^*_{\alpha=1})|}\cdot{c_\text{LTE}}.
    \label{eq:p_dev}
\end{equation}
Note that this measures the difference in behavior between w.r.t. to optimal policy but does not necessarily reflect a proportional decrease in performance. Instead, this is an measure of the normalized collective error, as in common estimators that try to project the optimal LTE-usage and system lifetime.

\item \emph{Relative reward loss:} Defines the relative loss in the expected total reward $\overline{R}(\bm{\pi})$ with policy $\bm{\pi}$ w.r.t. $\bm{\pi}^*_{\alpha=1}$ as
\begin{equation}
    \mathcal{L}(\bm{\pi})=\frac{\left|\overline{R}(\bm{\pi}^*_{\alpha=1})-\overline{R}(\bm{\pi})\right|}{\overline{R}(\bm{\pi}^*_{\alpha=1})}
\end{equation}

\end{itemize}

The results with the fully observable MDP were obtained analytically. In order to evaluate the performance of the POMDP and forgetful methods, analytical results were obtained for extreme values of parameter $\eta$.   For all other cases, we performed Monte-Carlo simulations of 20000 episodes. The duration of each episode depends on the system lifetime which could last up to several million time steps. 
\section{Results}
\label{sec:results}
In this section we investigate the performance of the modelled system. The investigation in this section is guided by the use of interface diversity in the case a combination of Wi-FI and LTE. The performance of the aforementioned system where $i=1$ is LTE and $i=2$ is Wi-Fi was evaluated by a Monte-Carlo Matlab simulation (when necessary) where the calculation of the statistical properties for the GE model are derived from experimental latency measurements. 

We tested the system for all 5 different values $\eta = 0, 0.03, 0.07, 0.2, 1$ where the fully observable MDP system had different transmission policies. Given the measurements and the characteristics of our measurement setup, we tuned the simulation parameters to the values in Table \ref{tab:setting}.

\begin{figure*}[t]
	\centering
	\includegraphics[width=0.75\linewidth]{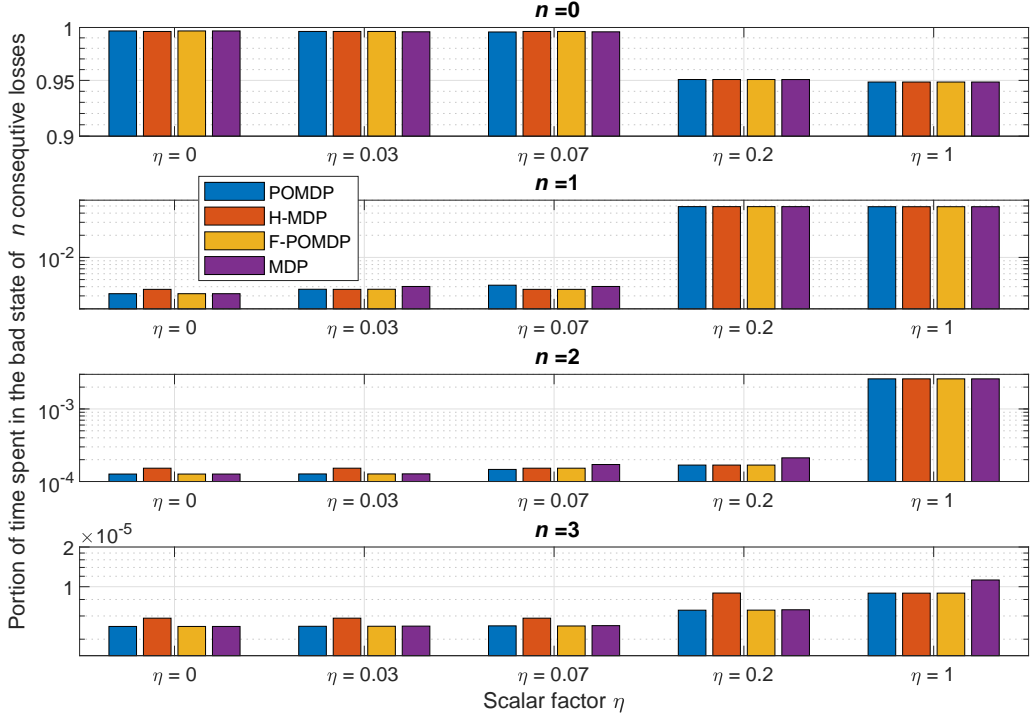}
	\caption{Portion of time spent in state with $n$ consecutive untimely packets for all simulated and analytically extracted data.
	}
	\label{fig:nrez}
\end{figure*}
\subsection{Extreme Policies}
\label{sec:cpol}
As a starting point, we describe and evaluate the policies obtained in the cases where the value of parameter $\eta\in\left(-\infty,\infty\right)$ is set to a extremely low or high value (i.e., at either of the extremes of its range).
When adjusting the scaling factor to its lowest possible value $\eta \to -\infty$, the cost of using each interface is omitted and the MDP optimizes in favor of not losing any transmissions. Thus, we observe an extreme behavior that is not affected by the belief, or the POMDP behavior. Specifically, the optimal policy maintains both interfaces turned on no matter the current state $\pi^*(S)=A(1,1) \, \forall S \in \mathcal{S}$. Since the utilization of both interfaces is 100\%, this gives the upper bound on burst error performance for the whole system. Even in this case, there is a non-zero probability to end up in the absorbing state which happens with an expectation of $5.0738 \cdot 10^6$ transitions. This is the defined lifetime of the system, and before turning-off due to failure the system maintains successful transmissions for $99.67\%$ of the time, $0.32\%$ of transmissions have a single error, $0.0127\%$ have two consecutive errors, and $4.9971 \cdot 10^{-4}\%$ of all burst errors have three consecutive errors. 

On the other hand, setting $\eta\to\infty$ creates a lower bound of the system performance that aims to minimize the cost of operation at the expense of decreasing the system lifetime. This is the result of scaling the cost of performing a transmission to be higher than the reward of maintaining successful transmissions. Since the action space is restricted to use at least one interface for transmission at all times, in such a cost restraint system, it is reasonable to only allow for the utilization of the Wi-Fi interface $\pi^*(S)=A(0,1) \, \forall S \in \mathcal{S}$, due to the high cost of using LTE. Since we have 100\% utilization of Wi-Fi and 0\% utilization of LTE the system lifetime of the system decreases drastically to $1.3633 \cdot 10^5$. During operation, the system maintains successful transmissions for $94.84\%$ of the time, $4.89\%$ of transmissions have a single error, $0.26\%$ have two consecutive errors, and $0.0138\%$ of all burst errors have three consecutive errors. All the implementations with other values of $\eta$ result in policies that exploit mixtures of actions and could not be analytically extracted -- for the POMDP and H-MDP implementations -- and are thus extracted through Monte-Carlo simulations, as detailed in the previous section.

\subsection{Optimal Policies with Scaled Costs}

\label{sec:opt_policies}

The portion of time spent in states with $n$ consecutive untimely packets, obtained from the simulations, are shown in Fig. \ref{fig:nrez} as a function of $\eta$, from which we can extract several conclusions.
Initially, we notice a sharp decay for the portion of time spent in good states when comparing the values with $\eta = 0.07$ and $\eta = 0.2$. In accord, we notice a sharp increase increase in all bad states, that is most significant for single errors. This manifests in the optimal policy, as a reluctance of mitigating single burst errors ($n=1$). Notwithstanding this increase in single errors, all approaches still mitigate higher orders of error bursts ($n>1$) when $\eta = 0.2$. This is not true for $\eta = 1$, where all approaches focus solely on mitigating the last error that may lead to exceeding the survival time $N$. 

It is important to notice that due to the fact that H-MDP treats the GE model as hidden when a portion of it is unobservable, turning off an interface results in fully losing the state for that interface. This leads to a behavior where the H-MDP would intentionally turn off the interface that observes a bad state, even when there is no negative incentive to keeping that interface on, in favor of the more likely transition to the steady state of the good state for that interface. Due to this, the optimal policy for H-MDP is never $\pi^*(S)=A(1,1) \, \forall S \in \mathcal{S}$, even when $\eta=0$. Interestingly, H-MDP uses the same policy for all three $\eta=0, 0.03, 0.07$, but is best fit for $\eta=0.07$. This makes H-MDP the only suboptimal approach -- out of all four -- for $\eta=0$.

\begin{figure}[t!]
	\centering
	\includegraphics[width=1\linewidth]{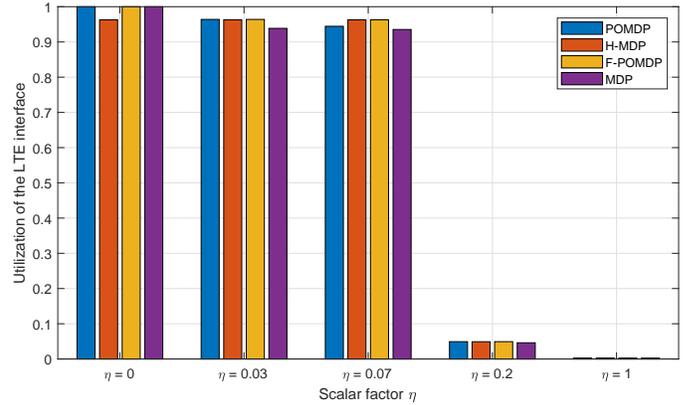}
	\caption{Portion of time spent using the LTE interface $i=1$.
	}
	\label{fig:iuse}
\end{figure}

\begin{figure}[tb]
	\centering
	\includegraphics[width=1\linewidth]{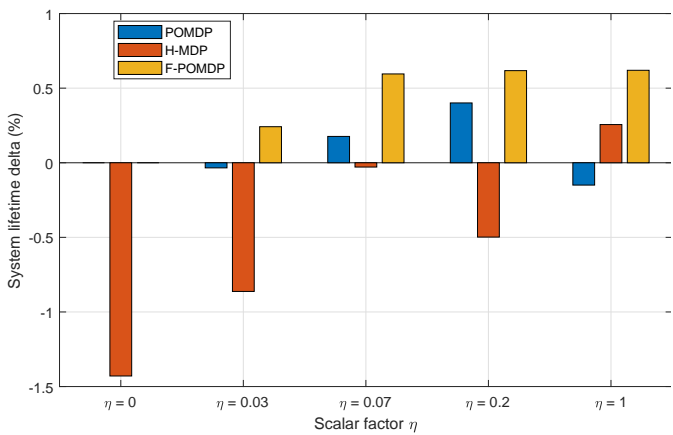}
	\caption{Relative increase in the system lifetime $\Delta K$ (i.e., in the time to reach one of the absorbing states with $N$ consecutive errors) w.r.t. to the fully observable, optimal policy of MDP .
	}
	\label{fig:sruvt}
\end{figure}

Due to the low cost of using the Wi-Fi interface, and generally superior $r$ probability, all policies keep the Wi-Fi interface at 100\% utilization. On the other hand, LTE utilization is the only that varies for each approach, and is shown in Fig \ref{fig:iuse}.
However, we are interested in investigating the lifetime of the system given burst error tolerance of $N$. The system lifetime for the different scaling factors $\eta$ is given in Fig.~\ref{fig:sruvt} as a difference from the optimal system lifetime. Therefore, the goal of each approach with limited observability is to follow the performance of the optimal, fully observable, approach as closely as possible. Thus, when a policy improves the system-lifetime, it is a sign of energy-inefficiency that comes in the form of extra LTE-interface utilization.
Accordingly, the goal of all three approaches that have to work with limited information is to achieve $\Delta K\approx 0$. Looking at Fig.~\ref{fig:sruvt}, we can also notice that aside from the case for $\eta=0.07$, the POMDP approach gives the least deviations with regards to the other two approaches. Moreover, the good performance of the H-MDP approach in the case for $\eta=0.07$ is a simple coincidence since this approach applies exactly the same policy for $0 \leq \eta \leq 0.07$, where both the POMDP and the F-POMDP tend to vary and adapt. Additionally, we notice that the F-POMDP approach is highly focused towards increasing the system lifetime which, as shown in Fig.~\ref{fig:iuse}, comes at the cost of using the LTE more often than with of the optimal approach. Since this behavior is quite consistent, we can safely say that the F-POMDP is a system-lifetime conservative approach. The POMDP approach is however more adaptable, and consistently bests the F-POMDP in replicating the system lifetime of the fully observable MDP.

\begin{figure}[tb]
	\centering
	\includegraphics[width=1\linewidth]{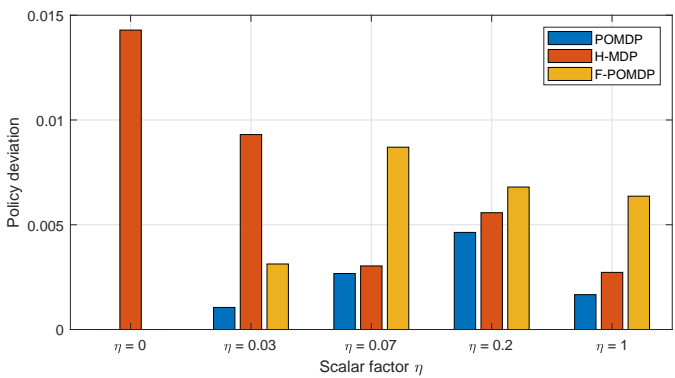}
	\caption{The deviation from the optimal MDP policy of full observably.}
	\label{fig:deviations}
\end{figure}

Finally, in Fig.~\ref{fig:deviations} we show the aggregate deviation, in terms of system lifetime and energy, as calculated as in~\eqref{eq:p_dev}. Here we can see that the POMDP approach provides the most-adaptable behavior, that best resembles the policy when having full observability. Treating the system as a hidden MDP does yield some adaptability, however, the approach can lead to large deviations from the optimal behavior, as it can be seen for $\eta = 0, 0.03$. In these cases, since the stochastic process was treated as hidden to the MDP, the H-MDP optimal solution would intentionally turn off the LTE interface when it is in the bad state. With this, H-MDP fully loses the information of the LTE interface, in favor of the better stationary state probabilities. Due to this, we consider the H-MDP approach as unsuitable. On the other hand, the F-POMDP was always conservative with regards to the system-lifetime. Thus, F-POMDP approach is ideal for implementation in devices with extreme power limitations, as it does not require re-computation of the belief states continuously. Finally, the POMDP approach provides the best solutions that most closely follow the optimal policy. Hence, it presents the best solution given that an accurate model of the environment is available and should be adopted if the energy consumption of the computational circuit, that is dedicated for updating the belief states, is not an issue. In the following, we present a sensitivity analysis of the considered methods under an imperfect model of the environment.
\subsection{Sensitivity Analysis}
\begin{figure}[tb]
	\centering
	\includegraphics[width=1\linewidth]{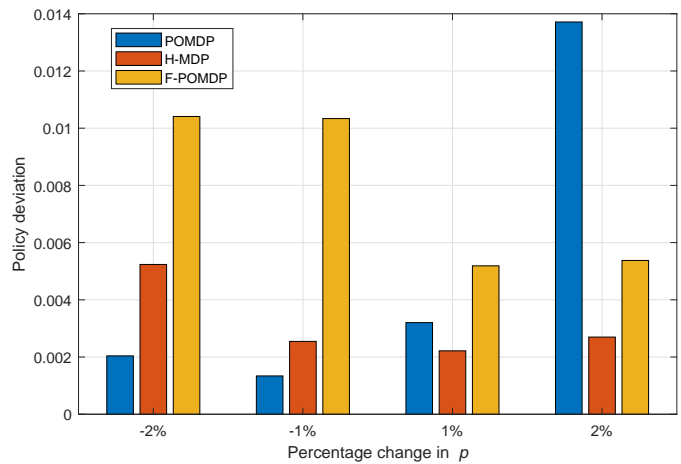}
	\caption{The deviation from the optimal MDP policy of full observably when having an error in estimating the $p$ and $r$ values, for the case of $\eta=0.07$.}
	\label{fig:stress}
\end{figure}
We conclude the section by evaluating the impact of the estimation error regarding the $p$ and $r$ values for $\eta=0.07$. We do this by adding a percentage of error to the $p$ value, while the $r$ is calculated to maintain the same steady state probabilities $\pi_{i,\B}$ and $\pi_{i,\G}$ for each interface $i$ as derived from the values in Table \ref{tab:setting}. In this way, the true probabilities of the Markov system are hidden from the decision processes. As Fig.~\ref{fig:stress} shows, in the case of a negative percentage change for the system, F-POMDP approach greatly deviates from the optimal policy for $\eta=0.07$. Additionally, the POMDP approach deviates considerably when an error of $2\%$ is introduced. Moreover, the H-MDP is the most robust as it is more reluctant to change policies in the presence of different parameters, which shows best in the case of positive errors. We can conclude that while belief mechanics help adapt to the optimal policy in the case where the model of the environment is perfectly known, however, such implementations can lead to bad results in particular scenarios where the true probabilities of the system are hidden from the agent. On the other hand, the H-MDP system does not show a big disadvantage in those cases since it already treats the Markov process as hidden. 

\begin{figure}[tb]
	\centering
	\includegraphics[width=1\linewidth]{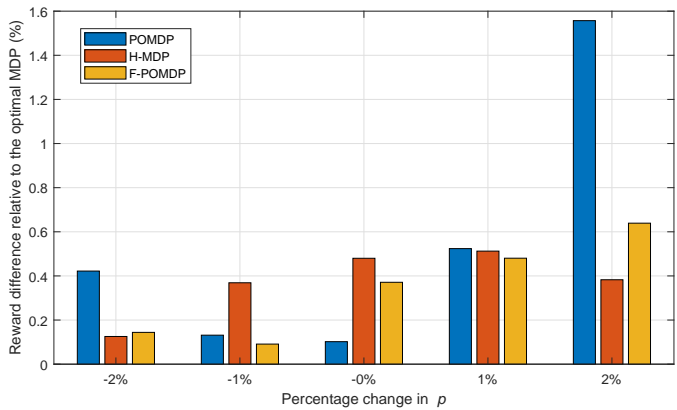}
	\caption{The system rewards with regards to the optimal MDP policy of full observably when having an error in estimating the $p$ and $r$ values, for the case of $\eta=0.07$.}
	\label{fig:stressrew}
\end{figure}

To conclude, we show the relative reward loss $\mathcal{L}(\bm{\pi})$ for the same cases of estimation error, as relative to the optimal MDP policy, in Fig.~\ref{fig:stressrew}. Here we observe that, even though the F-POMDP method deviates considerably from the optimal MDP policy in the negative estimations of -1\% -2\% (see Fig.~\ref{fig:stress}), its rewards are close to those with the optimal policy. The reward loss with the H-MDP are relatively stable and do not exceed $0.6$\%. Finally, we see that the POMDP implementation generally achieves a small reward loss, which is around $0.1$\% for no error and around $0.4$\% for -2\%. Nevertheless, a high loss is achieved with 2\% error. In this case, the adaptability of the POMDP has a negative effect since it scales the policy in accord with the erroneous $p$ and $r$ values. 

\section{Conclusion}
\label{sec:conclusion}
Motivated by the recent requirements for cyber-physical systems, we analyzed the problem of addressing error bursts by using two different wireless interfaces. We model the problem as a Gilbert-Elliott model with good and bad states for each interface. Given limited energy resources on our device, we derived and evaluated transmission policies to achieve an adequate trade-off between system lifetime and energy consumption with limited channel information. For this reason, we modelled the system as a POMDP that memorizes and calculates its belief for the observable states. Using Value Iteration to extract the Q-values from the MDP, we update the policy for the POMDP through the $Q_\text{MDP}$ technique. Our results show that the POMDP approach indeed produces near-optimal policies when the environment is accurately characterized. As such, this is a computationally inexpensive solution that closely follows the performance of the optimal policy, even in cases with various and mixed state-action pairs. We also propose a forgetful F-POMDP approach with only two finite belief states. This approach performs worse than the classic POMDP, with affinity to increase system-lifetime, but is well suited for approaches that are under extreme energy limitations. Finally, in future works we would like to practically validate the usefulness of the system onto several application scenarios, and address dynamic systems in non-stationary or non-characterized environments.

\section*{Conflict of Interest Statement}

The authors declare that the research was conducted in the absence of any commercial or financial relationships that could be construed as a potential conflict of interest.

\section*{Author Contributions}
ID: formal analysis and software development;
ID,ILM,JJN: conceptualization, investigation, writing; ID,ILM,JJN,PP: review, editing; PP,JJN: resources, funding acquisition, supervision, and project administration.

\section*{Funding}
The work was supported by the European Union's research and innovation programme under the Marie Sklodowska-Curie grant agreement No. 812991 ''PAINLESS'' within the Horizon 2020 Program.

\bibliographystyle{IEEEtran}
\bibliography{./bibliography.bib}

\end{document}

%% file: figures/gilbert_elliott_4state.tex
\begin{tikzpicture}[->, >=stealth', auto, semithick, state/.append style={minimum size=30pt, font=\normalsize, inner sep=1pt}]
\pgfmathsetmacro{\xshft}{4.6}
\pgfmathsetmacro{\yshft}{-1.4}
\pgfmathsetmacro{\axshft}{2}
\pgfmathsetmacro{\ayshft}{-1.2}
\node[state]    (A)[fill=Greens-C] at(\yshft,0)  {G};
\node[state]    (B)[ fill=Reds-E] at(-\yshft,0)  {B};
\path
 (A) edge[bend left=10]     node{$p_i$}         (B)
 (A) edge[in=200,out=160,loop]	    node[left]{$1-p_i$}         (A)
 (B) edge[bend left=10]     node{$r_i$}         (A)
 (B) edge[in=20,out=340,loop]	    node[right]{$1-r_i$}         (B);
 
          \node at (0,2.5*\yshft){(a)};

\begin{scope}[xshift= 7cm]

 \node[state]    (GG)[fill=Greens-C] at (0,0)  {G,G};
\node[state]    (GB)[ fill=Blues-B] at (-\yshft,-\yshft)  {G,B};
\node[state]    (BG)[ fill=Blues-B] at (-\yshft,\yshft)   {B,G};
\node[state]    (BB)[fill=Reds-E] at (-2*\yshft,0)   {B,B};
\path
 (GG) edge[bend left=10]              (GB)
 (GB) edge[bend left=10]              (GG)
 
 (GG) edge[bend left=10]              (BG)
 (BG) edge[bend left=10]              (GG)
 
 (GG) edge[bend left=10]              (BB)
 (BB) edge[bend left=10]              (GG)
 
 (GB) edge[bend left=10]              (BG)
 (BG) edge[bend left=10]              (GB)
 
 (GB) edge[bend left=10]              (BB)
 (BB) edge[bend left=10]              (GB)
 
 (BG) edge[bend left=10]              (BB)
 (BB) edge[bend left=10]              (BG)
 
  (GG) edge[in=200,out=160,loop]	             (GG)
 (GB) edge[in=110,out=70,loop]	             (GB)
  (BG) edge[in=290,out=250,loop]	             (BG)
 (BB) edge[in=20,out=-20,loop]	  
          (BB);
          
\node at (-\yshft,2.5*\yshft){(b)};
\end{scope}
\end{tikzpicture}

%% file: figures/pomdp.tex
\begin{tikzpicture}[->, >=stealth', auto, semithick, node distance=2.5cm, state/.append style={minimum size=35pt, font=\footnotesize, inner sep=1pt}]
\pgfmathsetmacro{\xshft}{4.6}
\pgfmathsetmacro{\yshft}{-1.4}
\pgfmathsetmacro{\axshft}{2}
\pgfmathsetmacro{\ayshft}{-1.2}
\node[state] (s) at (0,0)[ fill=Blues-B, align=center, label={above:State $S$}]  {$s_1,s_2,$\\$n<N$};
\node[state] (s1) at (\xshft,-2.5*\yshft)[fill=Greens-C, align=center]  {G,G,\\$0$};
\node[state] (s2) at (\xshft,-1.5*\yshft)[ fill=Greens-C, align=center]  {B,G,\\ $0$};
\node[state] (s3) at (\xshft,-0.5*\yshft)[ fill=Greens-C, align=center]  {G,B,\\ $0$};
\node[state] (s4) at (\xshft,0.5*\yshft)[fill=Reds-E, align=center] {B,B,\\$n+1$};
\node[state] (s5) at (\xshft,1.5*\yshft)[fill=Reds-E, align=center] {B,G,\\$n+1$};
\node[state] (s6) at (\xshft,2.5*\yshft)[fill=Reds-E, align=center] {G,B,\\$n+1$};

\coordinate (a1) at (\axshft,-\ayshft);
\coordinate (a2) at (\axshft,0);
\coordinate (a3) at (\axshft,\ayshft);

\filldraw[fill=Greys-D,Greys-D] (\axshft-1.2,-2*\yshft) rectangle (\axshft+0.5,2*\yshft);
\draw[thick] (\xshft+\axshft-0.2,-0.5*\yshft+0.4) rectangle (\xshft+1.5*\axshft +6,0.5*\yshft-0.4);

\node at (a1)[above left, inner sep=2pt]{$(1,1)$};
\node at (a2)[above left, inner sep=2pt]{$(1,0)$};
\node at (a3)[below left, inner sep=2pt]{$(0,1)$};
\path[Blues-K, thick]
 (s) edge[-, bend left=10] (a1)
 (a1) edge[bend left=15] (s1.190)
 (a1) edge[bend left=10] (s2)
 (a1) edge[bend right=10] (s3.170)
 (a1) edge[bend right=15] (s4)
  (\xshft+\axshft,-0.5*\yshft) edge node[right, pos=1, black]{Fully observable $O=S'$} (\xshft+1.5*\axshft,-0.5*\yshft);
    
  \path[Set2-A, thick]  
 (s) edge[-] (a2)
 (a2) edge[bend left=15] (s1.220)
 (a2) edge[bend left=10] (s3.200)
 (a2) edge[bend right=10] (s4)
 (a2) edge[bend right=15] (s5)
  (\xshft+\axshft,0) edge node[right, pos=1, black]{Partially observable with $o_2=0$} (\xshft+1.5*\axshft,0);
         
 \path[Set2-B, thick]
 (s) edge[-, bend right=10] (a3)
 (a3) edge[bend left=15] (s1.245)
 (a3) edge[bend left=10] (s2)
 (a3) edge[bend right=10] (s4)
 (a3) edge[bend right=15] (s6)
 (\xshft+\axshft,0.5*\yshft) edge node[right, pos=1, black]{Partially observable with $o_1=0$}  (\xshft+1.5*\axshft,0.5*\yshft)
   ;

\filldraw[fill=black] (a1) circle [radius=1.5pt];
\filldraw[fill=black] (a2) circle [radius=1.5pt];
\filldraw[fill=black] (a3) circle [radius=1.5pt];

\node at (\axshft-0.35,-2*\yshft)[above]{Action space $\mathcal{A}$};
\node at (\xshft,-3*\yshft)[above]{Next state $S'$};

\end{tikzpicture}